\documentclass{emulateapj}

\graphicspath{{./Figures/}}

\shorttitle{The nature of the hyper-runaway candidate \mbox{HIP\,60350}}
\shortauthors{Irrgang et al.}

\begin{document}

\title{The nature of the hyper-runaway candidate \mbox{HIP\,60350}\altaffilmark{1,2}}

\author{Andreas Irrgang\altaffilmark{3}, Norbert Przybilla\altaffilmark{3}, Ulrich Heber\altaffilmark{3}, M. Fernanda Nieva\altaffilmark{4,5} and Sonja Schuh\altaffilmark{6,7}}

\altaffiltext{1}{Based on observations obtained with the Hobby-Eberly Telescope, which is a joint project of the University of Texas at Austin, the Pennsylvania State University, Stanford University, Ludwig-Maximilians-Universit\"at M\"unchen, and Georg-August-Universit\"at G\"ottingen (proposal G09-1-002).}
\altaffiltext{2}{Based on observations collected at the Centro Astron\'{o}mico Hispano Alem\'{a}n (CAHA) at Calar Alto, operated jointly by the Max-Planck Institut f\"ur Astronomie and the Instituto de Astrof\'{i}sica de Andaluc\'{i}a (CSIC) (proposals H2005-2.2-016, F2009-3.5-008 and H2009-3.5-028).}
\altaffiltext{3}{Dr.~Karl Remeis Observatory \& ECAP, University of Erlangen-Nuremberg, Sternwartstrasse~7, D-96049~Bamberg, Germany; andreas.irrgang@sternwarte.uni-erlangen.de.}
\altaffiltext{4}{Visiting scientist at the Dr.~Karl Remeis Observatory.}
\altaffiltext{5}{Max Planck Institute for Astrophysics, Karl-Schwarzschild-Strasse~1, D-85741~Garching, Germany.}
\altaffiltext{6}{Georg-August-Universit\"at~G\"ottingen, Institute for Astrophysics, Friedrich-Hund-Platz~1, 37077~G\"ottingen, Germany.}
\altaffiltext{7}{TEA Visiting Professor, Institut f\"ur Astronomie und Astrophysik, Kepler Center for Astro and Particle Physics, Eberhard-Karls-Universit\"at, Sand~1, 72076~T\"ubingen.}

\begin{abstract}
Young, massive stars in the Galactic halo are widely supposed to be the result of an ejection event from the Galactic disk forcing some stars to leave their place of birth as so-called runaway stars. Here, we present a detailed spectroscopic and kinematic analysis of the runaway B-star \mbox{HIP\,60350} to determine which runaway scenario -- a supernova explosion disrupting a binary system or dynamical interaction in star clusters -- may be responsible for \mbox{HIP\,60350}'s peculiar orbit. Based on a non-local thermodynamic equilibrium approach, a high-resolution optical echelle spectrum was examined to revise spectroscopic quantities and for the first time to perform a differential chemical abundance analysis with respect to the B-type star \mbox{18\,Peg}. The results together with proper motions from the \textit{Hipparcos} Catalog further allowed the  three-dimensional kinematics of the star to be studied numerically. The abundances derived for \mbox{HIP\,60350} are consistent with a slightly supersolar metallicity agreeing with the kinematically predicted place of birth \mbox{$\sim$6\,kpc} away from the Galactic center. However, they do not exclude the possibility of an $\alpha$-enhanced abundance pattern expected in the case of the supernova scenario. Its outstanding high Galactic rest frame velocity of \mbox{$530\pm35\rm\,km\,s^{-1}$} is a consequence of ejection in the direction of Galactic rotation and slightly exceeds the local Galactic escape velocity in a standard Galactic potential. Hence \mbox{HIP\,60350} may be unbound to the Galaxy.
\end{abstract}


\section{Introduction}
Young, massive stars at high galactic latitudes are rare objects albeit known for decades \citep{blbain15}. According to \citet{tobin87}, there are several plausible alternatives to explain them. The possibilities encompass I) \textit{in situ} star formation caused by the collision of intermediate or high-velocity H\,\textsc{i} clouds, II) misinterpretation of rather hot evolved stars closely mimicking massive ones or III) the formation in the disk and subsequent ejection into the halo as so-called runaway stars. Possible ejection scenarios for the latter are dynamical interaction in star clusters -- either initial dynamical relaxation \citep{pobdlotyt4} or many-body encounters \citep{leaj96} -- or a supernova explosion in a binary system \citep{blbain15}.

Interest in runaway stars has been revived recently by the discovery of a new class of extreme velocity stars \citep{brapjl622,edapjl634,hiaap444}, the so-called hypervelocity stars (HVSs), traveling at such a high velocity that they escape from the Galaxy. They were first predicted by theory \citep{hinat331} to be the result of the tidal disruption of a binary system by a supermassive black hole (SMBH) that accelerates one component to beyond the Galactic escape velocity (the Hills mechanism). Because the Galactic center hosts such a SMBH it is the suggested place of origin for HVSs. However, the SMBH paradigm has been challenged recently by the young HVS \mbox{HD\,271791} because its kinematics point to a birthplace in the metal-poor rim of the Galactic disk \citep{heaap483}. \citet{prapj684} presented a high-precision quantitative spectral analysis and conclude that \mbox{HD\,271791} is the surviving secondary of a massive binary system disrupted in a supernova explosion. A similar scenario has been proposed for the origin of runaway B stars by \citet{blbain15}; hence, \citeauthor{prapj684} coined the term hyper-runaway star for the most extreme runaways that exceed the Galactic escape velocity.

\begin{table*}[t]
\centering
\caption{Stellar parameters and elemental abundances of \mbox{HIP\,60350} and \mbox{18 Peg}}
\label{stellar}
\begin{tabular}{l c c | l c c}     
\tableline\tableline
\multicolumn{3}{c|}{Stellar parameters}				&\multicolumn{3}{c}{$\log(\rm X/\rm H)+12$}\\
\tableline
		 						& \mbox{HIP\,60350} & \mbox{18 Peg}	 &	Element	& \mbox{HIP\,60350}	& \mbox{18 Peg}	\\
\tableline                
$T_{\rm eff}$\,[K]				& $16\,100\pm500$	& $15\,800\pm200$&	He		& $11.21$			& $10.99$		\\  
$\log (g\, \rm [cm\,s^{-2}])$	& $4.10\pm0.15$		& $3.75\pm0.05$  &	C		& $8.79\pm0.20$		& $8.33\pm0.09$	\\ 
$\xi\,\rm [km\,s^{-1}]$			& $5\pm2$			& $4\pm1$		 &	N		& $8.20\pm0.30$		& $7.80\pm0.11$	\\
$v \sin(i)\,\rm [km\,s^{-1}]$	& $150\pm5$			& $15\pm3$		 &	O		& $8.95\pm0.20$		& $8.80\pm0.11$	\\
$\zeta\,\rm [km\,s^{-1}]$		& $\ldots$			& $10\pm3$		 &	Mg		& $7.38\pm0.20$		& $7.51\pm0.07$ \\
$v_{\rm rad}\,\rm [km\,s^{-1}]$	& $262\pm5$			& $-14\pm1$		 &	Si		& $7.66\pm0.20$		& $7.51\pm0.11$ \\
$T_{\rm evol}$\,[Myr]			& $45^{+15}_{-30}$	& $61\pm5$		 &	S		& $7.38\pm0.20$		& $7.08\pm0.11$ \\
$d$\,[pc] 						& $3100\pm600$		& $362\pm21$	 &	Fe		& $7.18\pm0.25$		& $7.54\pm0.07$ \\
$M/M_{\sun}$					& $4.9\pm0.2$		& $5.6\pm0.2$	 &			& 					& \\
$R/R_{\sun}$					& $2.8\pm0.2$		& $5.2\pm0.3$	 &			& 					& \\
$\log (L/L_{\sun})$				& $3.3\pm0.6$		& $3.2\pm0.1$	 &			& 					& \\
\tableline
\end{tabular}
\tablecomments{The abundances are averages over all investigated lines of a chemical element. Since only one strong or a few weak lines were available for most elements when considering \mbox{HIP\,60350}, conservative abundance uncertainties were estimated by visual inspection (see Fig.~\ref{abundance_uncertainties}) instead of using the rms uncertainties from Table~\ref{linelist} as in the case of \mbox{18 Peg}. The macroturbulent velocity $\zeta$ is not measurable for \mbox{HIP\,60350} due to its large $v \sin(i)$.}
\end{table*}

\object{\mbox{HIP\,60350}} is a bright (\mbox{$V = 11.60\, \rm mag$}; \citealt{toaaps60}) mid-B-type star of high Galactic latitude $\left(l=144.6\degr, b=+75.1\degr\right)$. Among others, the star was extensively studied by \citet{maaap339} who estimated its atmospheric parameters from photometric indicators: effective temperature \mbox{$T_{\rm{eff}}=17\,100$\,K} and logarithmic surface gravity \mbox{$\log (g\, \rm [cm\,s^{-2}])=4.3$}, and they derived the radial velocity to \mbox{$v_{\rm{rad}}=217\pm20\,\rm km\,s^{-1}$}. Moreover, they inferred a distance of \mbox{$3.5$\,kpc}, a mass of \mbox{$5\,M_{\sun}$} and an age of about \mbox{$15\,\rm Myr$}. \citet{teaap369} used these values together with \textit{Hipparcos} proper motions to trace back the orbit of \mbox{HIP\,60350} to show that a dynamical disk runaway event \mbox{$20$\,Myr} ago is very likely the ejection mechanism in this case.

\mbox{HIP\,60350} has a total velocity referred to the local standard of rest (LSR) of \mbox{$v_{\rm{LSR}}=417\,\rm km\,s^{-1}$} \citep{maaap339} making it the second fastest runaway star after the massive B-type giant \mbox{HD\,271791}. The various similarities between both stars were motivation to us to re-investigate the origin of \mbox{HIP\,60350}.

To this aim we carried out a quantitative analysis of a high-resolution spectrum using non-local thermodynamic equilibrium (NLTE) techniques for the first time. Stellar parameters were thus revised and elemental abundances constrained (Sect.~\ref{spectroscopy}). The results together with proper motions from the new reduction of the \textit{Hipparcos} Catalog were used to determine the current three-dimensional (3D) space velocity. The following kinematic analysis suggested that the star originated in or near the Crux-Scutum spiral arm (Sect.~\ref{kinematics}). Finally we discuss the kinematic parameters and the elemental abundance pattern in the light of the rivaling formation scenarios, i.e., binary supernova versus dynamical cluster ejection (Sect.~\ref{discussion}).
\section{Spectroscopic Analysis}\label{spectroscopy}

\begin{figure*}[t]
   \centering
   \includegraphics[width=1\textwidth]{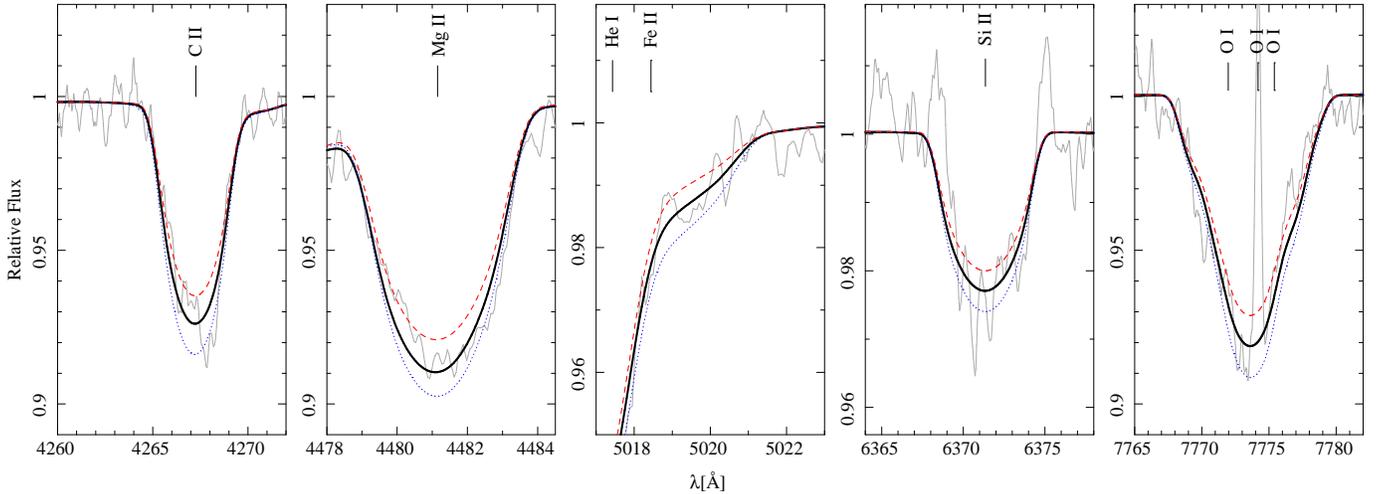}
   \caption{Determination of abundances and their uncertainties demonstrated in the cases of C, Mg, Fe, Si and O: the HET observation (gray solid line) is fitted best by the parameters of Table~\ref{stellar} (thick black line). The corresponding uncertainties are deduced visually by varying the abundance of an individual chemical species while keeping all others fixed. Blue dotted/red dashed lines stand for an increase/decrease of abundance according to the uncertainties in Table~\ref{stellar}.}
   \label{abundance_uncertainties}
\end{figure*}

\begin{table*}
\centering
\caption{Spectral line analysis of \mbox{HIP\,60350} and \mbox{18 Peg}}
\label{linelist}
\begin{tabular}{c c c c c c c}     
\tableline\tableline
$\lambda$(\AA)	&	$\chi(\rm{eV})$	&	$\log gf$	&	Accuracy	&	Source	&\multicolumn{2}{c}{$\log(\rm X/\rm H)+12$}	\\
\tableline
				&					&				&			&			&	\mbox{HIP\,60350}	& \mbox{18 Peg}		\\
\tableline
C\,\textsc{ii}:	&					&				&			&			&						&					\\
3918.98			&	16.33			&	-0.533		&	B		&	WFD		&		$\ldots$		&	8.19			\\
3920.69			&	16.33			&	-0.232		&	B		&	WFD		&		$\ldots$		&	8.30			\\
4267.00			&	18.05			&	0.563		&	C+		&	WFD		&		8.69			&	8.19			\\
4267.26			&	18.05			&	0.716		&	C+		&	WFD		&		-''-			&	-''-			\\
4267.26			&	18.05			&	-0.584		&	C+		&	WFD		&		-''-			&	-''-			\\
5132.94			&	20.70			&	-0.211		&	B		&	WFD		&		8.93			&	8.37			\\
5133.28			&	20.70			&	-0.178		&	B		&	WFD		&		-''-			&	-''-			\\
5145.16			&	20.71			&	0.189		&	B		&	WFD		&		8.73			&	8.34			\\
5151.09			&	20.71			&	-0.179		&	B		&	WFD		&		$\ldots$		&	8.39			\\
5662.47			&	20.71			&	-0.249		&	B		&	WFD		&		$\ldots$		&	8.41			\\
6578.05			&	14.45			&	-0.087		&	C+		&	N02		&		$\ldots$		&	8.25			\\
6582.88			&	14.45			&	-0.388		&	C+		&	N02		&		$\ldots$		&	8.37			\\
6783.90			&	20.71			&	0.304		&	B		&	WFD		&		$\ldots$		&	8.44		\\[1mm]
				&					&				&			&			&		\textbf{8.79}	&	\textbf{8.33}	\\
				&					&				&			&			&	$\pm$\textbf{0.13}	& $\pm$\textbf{0.09}\\
\tableline
\end{tabular}
\tablecomments{The table for all elements is available at the end of this paper. Accuracy indicators -- uncertainties within: A: 3\%; B: 10\%; C: 25\%; D: 50\%; E: larger than 50\%; X: unknown.
Sources of gf-values -- BB89: \citet{beaap209}; BMZ: \citet{bujopamp26}; FFT: \citet{fradandt87}; FFTI: \citet{fradandt92}; FMW: \citet{fu17}; FW: \citet{fu79}; KB: \citet{ku1995}; MAR: \citet{maaaps144}; MELZ: \citet{meadfoc28}; MERLR: \citet{majoqsart69}; N02: \citet{naadndt80}; N86: \citet{nuaap155}; WFD: \citet{wijpcrd7}; WSM: \citet{winsrds69}.}
\end{table*}

\mbox{HIP\,60350} was observed in 2008 December with the high-resolution echelle spectrograph of the $9.2$\,m Hobby-Eberly Telescope (HET) at the McDonald Observatory. Three individual spectra with resolving power $\lambda/\Delta \lambda=15\,000$ and useful wavelength range [3900\,\AA, 7870\,\AA] were co-added, resulting in a signal-to-noise ratio (S/N) around $140$ in the blue visual range. Additionally, three intermediate-resolution spectra taken in 2009 May and July with the $3.5$\,m telescope at Calar Alto, Spain, and its long-slit TWIN spectrograph were available enlarging the spectral coverage down to 3500\,\AA, making accessible the high-order Balmer lines and the Balmer jump.

The quantitative spectral analysis was carried out following the hybrid NLTE approach discussed by \citet{niapjl639, niaap467, niaap481} and \citet{praap445}: line-blanketed LTE model atmospheres were computed with ATLAS9 \citep{ku93}, while NLTE line formation calculations were performed using updated versions of DETAIL and SURFACE \citep{giphd, bu9}. State-of-the-art model atoms were adopted allowing absolute elemental abundances to be obtained with high accuracy (see \citealt{prapjl688,prapj684} for an overview). Atmospheric parameters and elemental abundances were derived by detailed line-profile analysis and fitting of the spectral energy distribution (SED).

\begin{figure*}[t]
   \centering
   \includegraphics[width=0.9\textwidth]{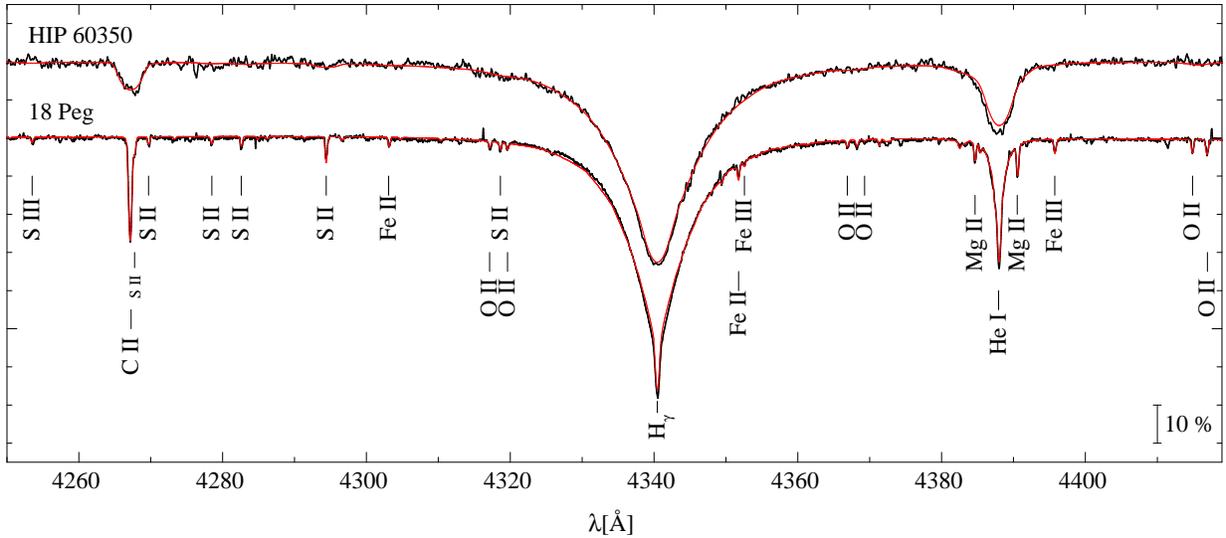}
   \caption{Comparison of final spectrum synthesis for \mbox{HIP\,60350}/\mbox{18\,Peg} (red line) with normalized HET/FOCES observation (black line) for an exemplary region around H$_{\gamma}$.}
   \label{comparison}
\end{figure*}

\begin{figure}[t]
   \centering
   \includegraphics[width=0.44\textwidth]{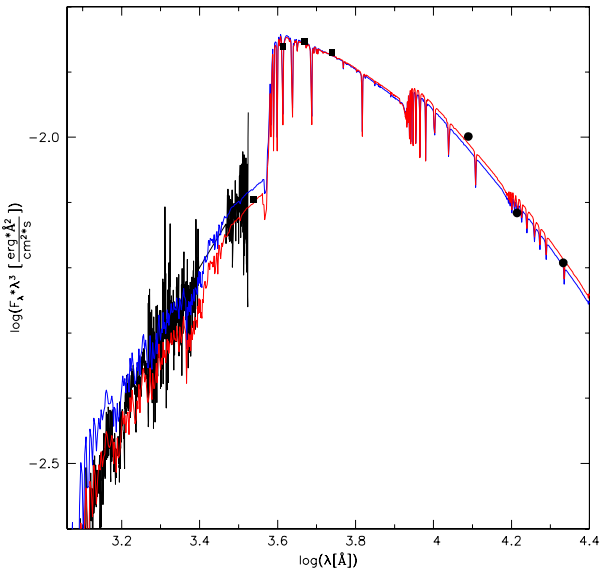}
   \caption{Observed SED from low-resolution IUE UV-spectro- (black line), $uvby$- (black filled squares; \citealp{toaaps60}) and $JHK$-photometry (black filled circles, \citealp{2mass}). The measurements are plotted vs.\ two different ATLAS9 models. The red one, calculated for the parameters of Table~\ref{stellar} (helium enriched), obviously fits better than the blue one, which is the best model assuming a standard solar helium abundance ($T_{\rm eff} = 16\,900\,\rm K$, $\log (g\, \rm [cm\,s^{-2}])=4.18$). Interstellar extinction was corrected by using a color excess of $E(B-V)=0.023\,\rm mag$ according to \citet{scapj500}.}
   \label{sed}
\end{figure}

The fundamental atmospheric parameters {$T_{\rm{eff}}=16\,100\pm500\,\rm K$ and $\log (g\, \rm [cm\,s^{-2}])=4.10\pm0.15$}, microturbulent velocity \mbox{$\xi=5\pm2\,\rm km\,s^{-1}$} and projected rotational velocity \mbox{$v \sin\,i = {150\pm5}\,\rm km\,s^{-1}$} were primarily constrained from Balmer and \mbox{He \textsc{i}} lines as well as the \mbox{Si \textsc{ii/iii}} ionization equilibrium. Elemental abundances were then obtained by matching the measurable lines of the individual chemical species while keeping all other stellar parameters fixed (see Fig.~\ref{abundance_uncertainties}). In the end, a final synthetic spectrum was computed which excellently reproduces the observation (see Fig.~\ref{comparison}), confirming the B-type nature of \mbox{HIP\,60350}. Interestingly, a helium abundance higher than solar, \mbox{$\log(\rm He/ \rm H)+12=11.21$}, was required to match the helium lines, the depth of the Balmer lines and the SED simultaneously (see Fig.~\ref{sed}). The resulting abundances (averaged over all lines of an element) are listed in Table~\ref{stellar}.

All spectra yielded a barycentric radial velocity of \mbox{$v_{\rm rad}=+262\pm5\,\rm km\,s^{-1}$}, equivalent to \mbox{$v_{\rm rad\_LSR}=+268\pm5\,\rm km\,s^{-1}$} in very good agreement with \citet{deaap202} who found \mbox{$v_{\rm rad\_LSR}=+270\,\rm km\,s^{-1}$}. Bearing in mind the different time intervals between each measurement, the star is unlikely a binary.

\begin{figure}[t]
   \centering
   \includegraphics[width=0.44\textwidth]{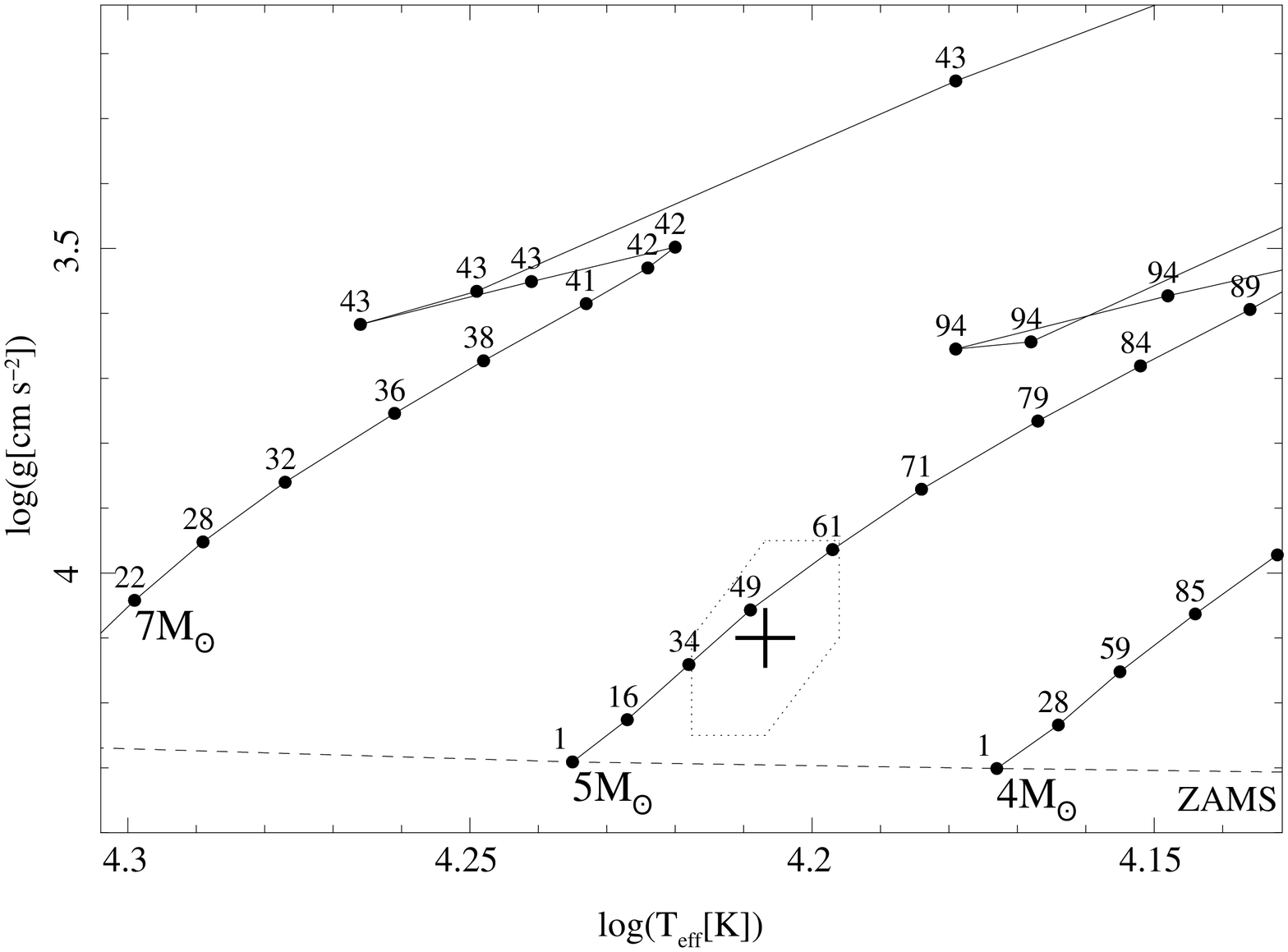}
   \caption{Position of \mbox{HIP\,60350} in a ($T_{\rm eff},\log g$) diagram with evolution tracks calculated by \citet{scaaps96}. Time steps are marked by filled circles giving the age in Myr. Uncertainties are indicated by the dotted hexagon. The locus of the ZAMS is displayed as well.}
   \label{evolution}
\end{figure}

Comparing the location of \mbox{HIP\,60350} in the $\left(T_{\rm eff},\log g \right)$ diagram to evolutionary tracks \citep{scaaps96} of solar metallicity as shown in Fig.~\ref{evolution} allowed its mass \mbox{$M=4.9\pm0.2\,M_{\sun}$} and age \mbox{$T_{\rm evol}=45^{+15}_{-30}\,\rm Myr$} to be constrained. The distance to \mbox{HIP\,60350} could then be calculated from $M$, $V$, $T_{\rm eff}$, $\log g$ and extinction $A_V=3.1\,E(B-V)=0.07\,\rm mag$ using the method described by \citet{raaap379} to be  \mbox{$d=3.1\pm0.6\,\rm kpc$}.

The precision of the analysis was restricted by an interplay of three effects: I) \mbox{HIP\,60350} lies in a temperature region where the optical spectrum shows very few strong but many weak metal lines. II) A considerable fraction of the latter is smeared out due to the high projected rotational velocity \mbox{$v \sin i$}. III) The S/N of the available spectra is often not sufficient to accurately determine the continuum behavior. Therefore, only few lines (in the case of magnesium and iron just one, see Table~\ref{linelist}) are above the noise level and thus usable.

\begin{figure}[t]
   \centering
   \includegraphics[width=0.44\textwidth]{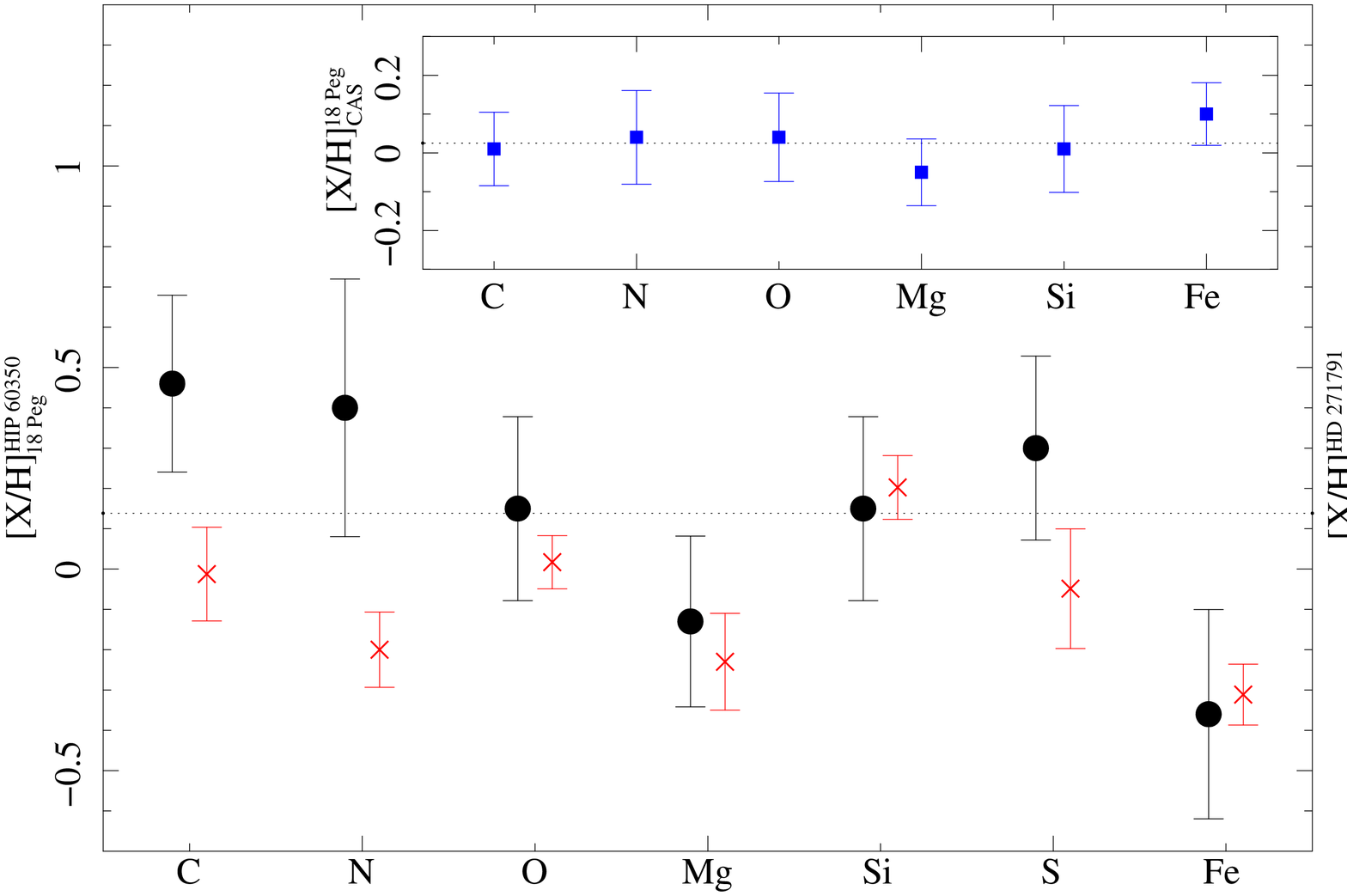}
   \caption{Metal abundances of \mbox{HIP\,60350} (black filled circles) and \mbox{HD\,271791} (red crosses, \citealp{prapj684}) relative to the reference star \mbox{18\,Peg}. We use the notation $\left[\rm X/\rm H\right]^{\rm{A}}_{\rm{B}}\equiv\log(\rm X/\rm H)_{\rm{A}}-\log(\rm X/\rm H)_{\rm{B}}$. The inset shows the abundances of \mbox{18\,Peg} (blue filled squares) relative to the CAS by \citet{prapjl688} revealing that \mbox{18\,Peg} is representative of the solar neighborhood. The dotted lines mark the average of the filled circles or squares, respectively.}
   \label{abundances}
\end{figure}

In order to exclude the possibility that the few lines measurable for \mbox{HIP\,60350} give abundances that are systematically higher or lower than other lines of the same chemical species, we also carried out an abundance analysis of the normal Population I B-type star \object{\mbox{18\,Peg}}. We used the Fibre Optics Cassegrain Echelle Spectrograph (FOCES; \citealt{pfaaps130}) on the 2.2 meter telescope at Calar Alto, Spain, to obtain a high-resolution, high-S/N spectrum ($\lambda/\Delta \lambda \approx 40\,000$, ${\rm{S/N}}\approx 400$ in the blue) of \mbox{18\,Peg}. Due to the high quality of the spectrum and the establishment of four ionization equilibria, atmospheric parameters and abundances could be constrained with high accuracy (see Fig.~\ref{comparison} for a comparison of the final synthetic spectrum with observation). This particular star was chosen since its parameters are similar to \mbox{HIP\,60350} (see Table~\ref{stellar}). Its chemical composition is representative of young B-type stars in the solar neighborhood, as can be seen from the comparison with the cosmic abundance standard (CAS) by \citet{prapjl688}; see the inset of Fig.~\ref{abundances}. The differential abundance pattern of \mbox{HIP\,60350} relative to the reference star \mbox{18\,Peg} is shown in Fig.~\ref{abundances}. The results obtained so far are summarized in Table~\ref{stellar}, while a detailed list of lines used for both spectroscopic investigations can be found in Table~\ref{linelist}. From the latter, we conclude that our quantitative analysis is not affected by a systematic bias.

\section{Kinematic Analysis}\label{kinematics}
\begin{figure*}
   \centering
   \includegraphics[width=1\textwidth]{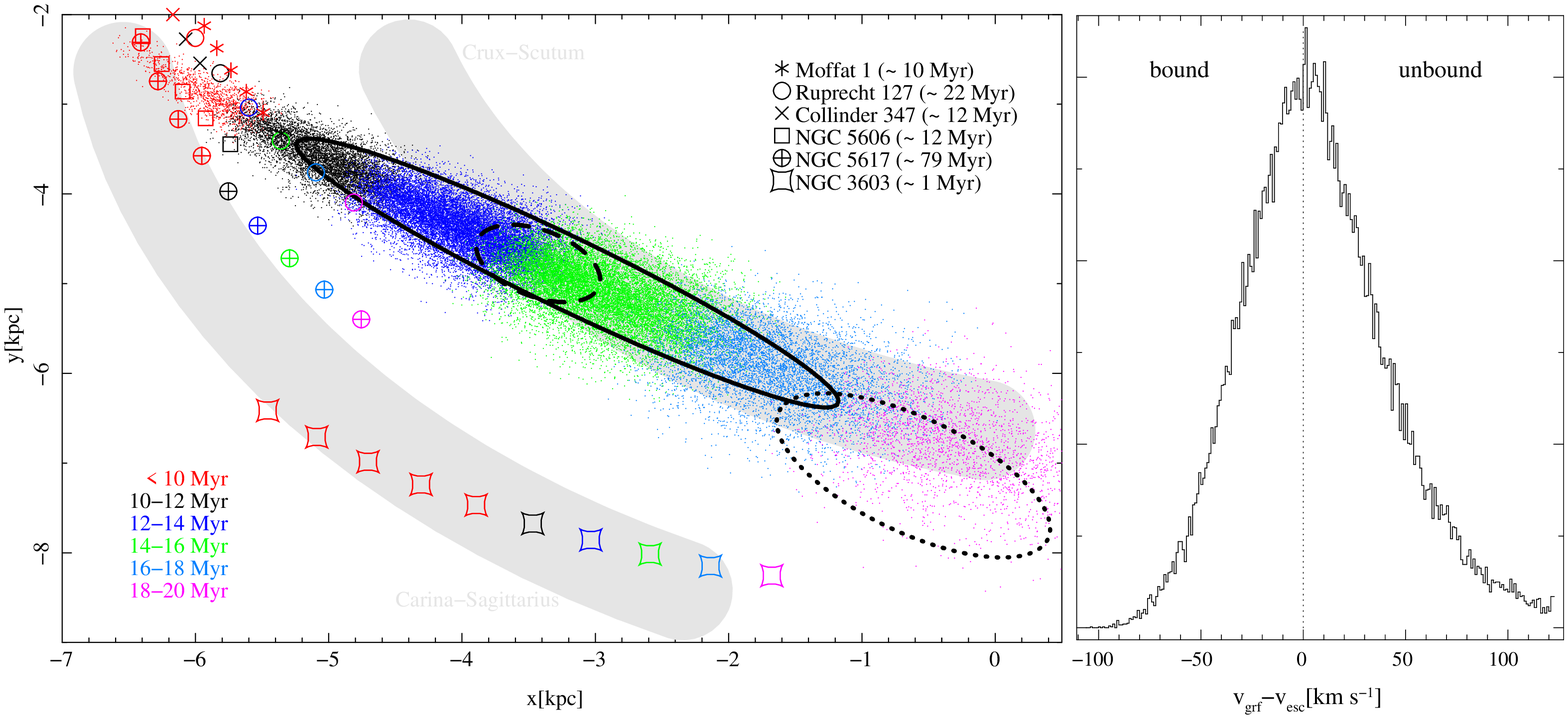}
   \caption{Left panel: Galactic disk intersection region of the sample of orbits (dots). Additionally, the trajectories of the best open cluster candidates are depicted using data from \citet{diaap389}. The flight time's color code is given in the lower left corner. A right-handed, non-rotating frame of reference with the Galactic center at the origin, Galactic north pole in the positive $z$-direction and the Sun's current position $\left(-8.0,0,0\right)$\,kpc is used. The ellipses mark the $1\sigma$ region with (solid) or without (dashed, dotted) accounting for uncertainties in distance, respectively, whereby the last one is the result of our computation based on the data from \citet{teaap369}. The gray-shaded regions schematically represent the locus of two spiral arms 14 Myr ago. Right panel: histogram showing the distribution of (un)bound trajectories in the sample. The abscissa is the difference between current space velocity $v_{\rm grf}$ and the local escape velocity $v_{\rm esc}$ according to the potential of Allen \& Santillan (1991).}
   \label{clusters}
\end{figure*}

The knowledge of proper motions provides access to the full 3D kinematics of \mbox{HIP\,60350} because distance and radial velocity are known (see above). In this work, we made use of the new reduction of the \textit{Hipparcos} Catalog \citep{hipparcos}, giving \mbox{$\mu_{\alpha}\cos(\delta)=-13.51\pm1.31\,\rm mas\,yr^{-1}$} and \mbox{$\mu_{\delta}=16.34\pm1.37\,\rm mas\,yr^{-1}$}. The Vizier database \citep{vizier} lists five ground-based measurements that are in excellent agreement with the \textit{Hipparcos} values. Consequently, the current space velocity relative to the LSR is \mbox{$v_{\rm{LSR}}=407\,\rm km\,s^{-1}$}. This value transforms into a total Galactic rest frame velocity of \mbox{$v_{\rm{grf}}=530\pm35\,\rm km\,s^{-1}$}, slightly exceeding the Milky Way's local escape velocity \mbox{$v_{\rm{esc}}=522\,\rm km\,s^{-1}$} derived from the Galactic gravitational potential of \citet{alrmxaa22}. Hence a detailed kinematic study seemed worthwhile.

To do so, the Galactic potential of \citeauthor{alrmxaa22} as well as the numerical code of \citet{odan313} was applied. This allowed the stellar orbit to be traced back  to the Galactic plane to investigate whether \mbox{HIP\,60350} is a runaway star at all and to determine the flight time $T_{\rm flight}$. Uncertainties were constrained via a Monte Carlo method that simultaneously and independently varied the initial parameters (i.e., the components of position and velocity), assuming for each a Gaussian distribution. The following results are average or -- when explicitly noted -- median values of a sample of 50\,000 trajectories. Uncertainties are expressed by the standard deviation. 

The obtained travel time \mbox{$T_{\rm flight}=14\pm3\,\rm Myr$} is shorter than the evolutionary timescale \mbox{$T_{\rm evol}=45^{+15}_{-30}\,\rm Myr$} and hence consistent with a runaway nature of the star\footnote{For comparison, \citet{teaap369} derive \mbox{$T_{\rm flight}=20.4\,\rm Myr$} while \citet{maaap339} give \mbox{$T_{\rm evol}\approx 15\,\rm Myr$}.}. To clarify the question whether \mbox{HIP\,60350} is bound to the Galaxy or not, we compared the current space and local escape velocity for each of our sample's orbits (see the right panel of Fig.~\ref{clusters}) showing that more than half of the trajectories (median value for $v_{\rm{grf}}-v_{\rm{esc}}$: \mbox{$5\,\rm km\,s^{-1}$}) computed for \mbox{HIP\,60350} are unbound to the Milky Way. Considering Galactic gravitational potentials with a lower \citep{xuapj684} or larger \citep{abapjl691} dark matter halo mass yields an overwhelming majority of unbound or bound orbits, respectively.

The disk intersection points of our sample of orbits are depicted in the left panel of Fig.~\ref{clusters} and imply an origin in the Galactic plane \mbox{$6.1\pm0.6\,\rm kpc$} away from the Galactic center, excluding a SMBH and hence the Hills mechanism as ejection scenario. The solid ellipse marks the $1\sigma$ region, i.e., it contains about 68\% of all points. The dashed $1\sigma$ ellipse is the result of a calculation without taking variations of position into account and is intended to show the dramatic effects of the distance uncertainty. The latter is to be compared to the dotted $1\sigma$ ellipse, which is our result when reproducing the conditions given in \citet{teaap369}; see Sect.\ \ref{dynintinstcl}. In addition, the figure schematically shows the locus of the two Galactic spiral arms Carina-Sagittarius and Crux-Scutum 14 Myr ago as gray-shaded areas, indicating that the star originated in or near the latter. Their positions were estimated from the polynomial logarithmic arm model of \citet{hoaap499} and the Galactic rotation curve of \citet{alrmxaa22}.

Our calculations also allowed the distribution of ejection velocities $v_{\rm ej}$ to be examined, defined as the absolute velocity at disk intersection relative to the rotating Galactic rest frame, yielding a median of \mbox{$379\,\rm km\,s^{-1}$} or an average of \mbox{$389\pm43\,\rm  km\,s^{-1}$}, respectively.
\section{Discussion}\label{discussion}
According to the results of Sect.\ \ref{kinematics}, \mbox{HIP\,60350} clearly qualifies as a candidate hyper-runaway star, i.e., an unbound runaway star. It has an ejection velocity similar to \mbox{HD\,271791} (\mbox{$v_{\rm ej}\approx 400\,\rm  km\,s^{-1}$}; \citealt{prapj684}), the first discovered hyper-runaway star, and was also expelled in the direction of Galactic rotation to reach its extreme current space velocity. Neither of the two ejection mechanisms -- binary supernova explosion and dynamical interaction -- can be strictly confirmed nor rejected for \mbox{HIP\,60350}.
\subsection{Supernova explosion in a binary system}
Various recent measurements of the Galactic metallicity gradient \citep[e.g.,][]{daapj617} show a large scatter of abundances for individual stars. Hence, the low iron abundance of Table~\ref{stellar} may not be in contradiction to the star's birthplace \mbox{$\sim$6\,kpc} away from the Galactic center and may in fact represent \mbox{HIP\,60350}'s baseline metallicity. On that condition, Fig.~\ref{abundances} reveals an enhancement of $\alpha$-elements resembling that of \mbox{HD\,271791} and thus giving indication for a supernova ejection event. Therefore, an analogous scenario as for \mbox{HD\,271791} can be devised (see \citealt{prapj684} for details). In brief, together with a primary star of at least \mbox{25\,$M_{\sun}$} on the zero-age main sequence (ZAMS), \mbox{HIP\,60350} formed a binary system that underwent a common-envelope phase. The resulting viscous forces triggered the spiraling in of the secondary component until the primary's shell was expelled due to the deposition of orbital energy by the secondary, leaving the primary behind as a Wolf-Rayet star. Eventually, the resulting very close binary system was disrupted by the primary's supernova explosion releasing \mbox{HIP\,60350} at almost orbital velocity. Considering a \mbox{15\,$M_{\sun}$} Wolf-Rayet star and making typical assumptions (synchronization of orbital and rotational periods/axes of the runaway star, circular orbits, symmetric supernova explosion to relate orbital and ejection velocity according to \citealt{taaap330}, \mbox{1.4\,$M_{\sun}$} supernova remnant) a pre-supernova system, with separation \mbox{$\sim$11$\,R_{\sun}$} and period \mbox{$\sim$0.9\,days} that is consistent with all observational constraints can be constructed. Requiring the Wolf-Rayet star to be of subclass WN could account for \mbox{HIP\,60350}'s high helium and nitrogen abundances. 
\subsection{Dynamical interaction in star clusters}\label{dynintinstcl}
Alternatively, Fig.~\ref{abundances} can be interpreted to show a metallicity slightly exceeding that of \mbox{18\,Peg} as well as the CAS. This is consistent with the kinematically predicted birthplace of Sect.\ \ref{kinematics}, in particular as \mbox{HIP\,60350's} iron abundance is deduced from a line which gives the lowest value for the reference star (see Table~\ref{linelist}). In that case, a dynamical ejection from a star cluster may be more likely. Knowing the flight time to the Galactic plane and the corresponding intersection region, we searched the catalog of open clusters by \citet{diaap389} for objects that match in position at the respective travel time as well as evolutionary age. For our star we found at least five appropriate candidates (see the left panel of \ Fig.~\ref{clusters}): Ruprecht 127 (age in Myr: $\sim$22), NGC\,5606 ($\sim$12), NGC\,5617 ($\sim$79), Collinder 347 ($\sim$12) and Moffat 1 ($\sim$10). The vicinity of the current massive cluster NGC\,3603 has been suggested as the possible spatial origin by \citet{teaap369}, but we find it not to be suited because its trajectory is far from the Galactic plane intersection area of \mbox{HIP\,60350}. Note that the catalog contains full 3D velocity information for the first three objects while all others were assumed to move on circular orbits around the Galactic center. It is important to mention that accounting for errors in the cluster trajectories analogously to the stellar orbit would noticeably increase the overlapping regions. Furthermore, many promising candidates, e.g., located in the Crux-Scutum spiral arm of the Milky Way, are very likely obscured by interstellar extinction and hence have not yet been discovered as already stated by \citet{teaap369}. The supersolar helium abundance of \mbox{HIP\,60350} might be the consequence of a merger event that occurred in the course of the dynamical ejection \citep[see][]{lemnras277}.
\section{Summary}
To summarize, based on a detailed spectroscopic and kinematic study we were able to significantly refine the stellar parameters, distance and possible place of birth of \mbox{HIP\,60350} revealing clear indications for a hyper-runaway nature of the star. Additionally, elemental abundances were derived via a NLTE analysis allowing the two competing ejection mechanisms of the runaway star, supernova explosion or dynamical interaction, to be discussed based on information previously not at hand. Performing this kind of investigation on optical spectra with a higher S/N than currently available might finally constrain \mbox{HIP\,60350}'s origin unambiguously.
\acknowledgments

We thank S.\ M\"uller and T.\ Kupfer for taking the TWIN spectra, H.\ Edelmann for support with the reduction of the HET data, H.\ Hirsch for providing access to his spectrum analyzing software SPAS and M.\ Firnstein for various pieces of advice, especially concerning the use of SPAS. Travel to the DSAZ (Calar Alto, Spain) was supported by the Deutsche Forschungsgemeinschaft (DFG) under grants He1356/50-1 and He1356/52-1. Publication costs were partly covered by DFG (grant He1356/45-1).
{\it Facilities:} \facility{HET (HRS)}, \facility{CAO:3.5m (TWIN)}, \facility{CAO:2.2m (FOCES)}, \facility{IUE}


\begin{thebibliography}{52}
\expandafter\ifx\csname natexlab\endcsname\relax\def\natexlab#1{#1}\fi

\bibitem[{{Abadi} {et~al.}(2009){Abadi}, {Navarro}, \& {Steinmetz}}]{abapjl691}
{Abadi}, M.~G., {Navarro}, J.~F., \& {Steinmetz}, M. 2009, \apjl, 691, L63

\bibitem[{{Allen} \& {Santillan}(1991)}]{alrmxaa22}
{Allen}, C., \& {Santillan}, A. 1991, RevMexAA, 22, 255

\bibitem[{{Becker} \& {Butler}(1989)}]{beaap209}
{Becker}, S.~R., \& {Butler}, K. 1989, \aap, 209, 244

\bibitem[{{Blaauw}(1961)}]{blbain15}
{Blaauw}, A. 1961, \bain, 15, 265

\bibitem[{{Brown} {et~al.}(2005){Brown}, {Geller}, {Kenyon}, \&
  {Kurtz}}]{brapjl622}
{Brown}, W.~R., {Geller}, M.~J., {Kenyon}, S.~J., \& {Kurtz}, M.~J. 2005,
  \apjl, 622, L33

\bibitem[{{Butler} \& {Giddings}(1985)}]{bu9}
{Butler}, K., \& {Giddings}, J.~R. 1985, in Newsletter of Analysis of
  Astronomical Spectra, No. 9 (London: Univ. London)

\bibitem[{{Butler} {et~al.}(1993){Butler}, {Mendoza}, \&
  {Zeippen}}]{bujopamp26}
{Butler}, K., {Mendoza}, C., \& {Zeippen}, C.~J. 1993, J.\ Phys.\ B, 26, 4409

\bibitem[{{Cutri} {et~al.}(2003){Cutri}, {Skrutskie}, {van Dyk}, {Beichman},
  {Carpenter}, {Chester}, {Cambresy}, {Evans}, {Fowler}, {Gizis}, {Howard},
  {Huchra}, {Jarrett}, {Kopan}, {Kirkpatrick}, {Light}, {Marsh}, {McCallon},
  {Schneider}, {Stiening}, {Sykes}, {Weinberg}, {Wheaton}, {Wheelock}, \&
  {Zacarias}}]{2mass}
{Cutri}, R.~M., {et~al.} 2003, 2MASS All Sky Catalog of Point Sources (The IRSA
  2MASS All-Sky Point Source Catalog, NASA/IPAC Infrared Science Archive),
  http://irsa.ipac.caltech.edu/applications/Gator/

\bibitem[{{Daflon} \& {Cunha}(2004)}]{daapj617}
{Daflon}, S., \& {Cunha}, K. 2004, \apj, 617, 1115

\bibitem[{{de Boer} {et~al.}(1988){de Boer}, {Richtler}, \& {Heber}}]{deaap202}
{de Boer}, K.~S., {Richtler}, T., \& {Heber}, U. 1988, \aap, 202, 113

\bibitem[{{Dias} {et~al.}(2002){Dias}, {Alessi}, {Moitinho}, \&
  {L{\'e}pine}}]{diaap389}
{Dias}, W.~S., {Alessi}, B.~S., {Moitinho}, A., \& {L{\'e}pine}, J.~R.~D. 2002,
  \aap, 389, 871

\bibitem[{{Edelmann} {et~al.}(2005){Edelmann}, {Napiwotzki}, {Heber},
  {Christlieb}, \& {Reimers}}]{edapjl634}
{Edelmann}, H., {Napiwotzki}, R., {Heber}, U., {Christlieb}, N., \& {Reimers},
  D. 2005, \apjl, 634, L181

\bibitem[{{Froese Fischer} \& {Tachiev}(2004)}]{fradandt87}
{Froese Fischer}, C., \& {Tachiev}, G. 2004, At.\ Data Nucl.\ Data Tables, 87,
  1

\bibitem[{{Froese Fischer} {et~al.}(2006){Froese Fischer}, {Tachiev}, \&
  {Irimia}}]{fradandt92}
{Froese Fischer}, C., {Tachiev}, G., \& {Irimia}, A. 2006, At.\ Data Nucl.\
  Data Tables, 92, 607

\bibitem[{{Fuhr} {et~al.}(1988){Fuhr}, {Martin}, \& {Wiese}}]{fu17}
{Fuhr}, J.~R., {Martin}, G.~A., \& {Wiese}, W.~L. 1988, J.\ Phys.\ Chem.\
  Ref.\ Data, Vol.\ 17, Suppl.\ 4 (New York: American Institute of Physics
and American Chemical Society)
 
\bibitem[{{Fuhr} \& {Wiese}(1998)}]{fu79}
{Fuhr}, J.~R., \& {Wiese}, W.~L. 1998, in CRC Handbook of Chemistry and
  Physics, ed. D.\ R.\ Lide ( 79th ed.; Boca Raton, FL: CRC Press)

\bibitem[{{Giddings}(1981)}]{giphd}
{Giddings}, J.~R. 1981, PhD thesis, Univ. London

\bibitem[{{Heber} {et~al.}(2008){Heber}, {Edelmann}, {Napiwotzki}, {Altmann},
  \& {Scholz}}]{heaap483}
{Heber}, U., {Edelmann}, H., {Napiwotzki}, R., {Altmann}, M., \& {Scholz},
  R.-D. 2008, \aap, 483, L21

\bibitem[{{Hills}(1988)}]{hinat331}
{Hills}, J.~G. 1988, \nat, 331, 687

\bibitem[{{Hirsch} {et~al.}(2005){Hirsch}, {Heber}, {O'Toole}, \&
  {Bresolin}}]{hiaap444}
{Hirsch}, H.~A., {Heber}, U., {O'Toole}, S.~J., \& {Bresolin}, F. 2005, \aap,
  444, L61

\bibitem[{{Hou} {et~al.}(2009){Hou}, {Han}, \& {Shi}}]{hoaap499}
{Hou}, L.~G., {Han}, J.~L., \& {Shi}, W.~B. 2009, \aap, 499, 473

\bibitem[{{Kurucz}(1993)}]{ku93}
{Kurucz}, R.~L. 1993, CD-ROM 13 (Cambridge: SAO)

\bibitem[{{Kurucz} \& {Bell}(1995)}]{ku1995}
{Kurucz}, R.~L., \& {Bell}, B. 1995, Kurucz CD-ROM No.\ 23 (Cambridge, MA:
  Smithsonian Astrophysical Observatory)

\bibitem[{{Leonard}(1995)}]{lemnras277}
{Leonard}, P.~J.~T. 1995, \mnras, 277, 1080

\bibitem[{{Leonard} \& {Duncan}(1988)}]{leaj96}
{Leonard}, P.~J.~T., \& {Duncan}, M.~J. 1988, \aj, 96, 222

\bibitem[{{Maitzen} {et~al.}(1998){Maitzen}, {Paunzen}, {Pressberger},
  {Slettebak}, \& {Wagner}}]{maaap339}
{Maitzen}, H.~M., {Paunzen}, E., {Pressberger}, R., {Slettebak}, A., \&
  {Wagner}, R.~M. 1998, \aap, 339, 782

\bibitem[{{Mar} {et~al.}(2000){Mar}, {P{\'e}rez}, {Gonz{\'a}lez}, {Gigosos},
  {del Val}, {de la Rosa}, \& {Aparicio}}]{maaaps144}
{Mar}, S., {P{\'e}rez}, C., {Gonz{\'a}lez}, V.~R., {Gigosos}, M.~A., {del Val},
  J.~A., {de la Rosa}, I., \& {Aparicio}, J.~A. 2000, \aaps, 144, 509

\bibitem[{{Matheron} {et~al.}(2001){Matheron}, {Escarguel}, {Redon}, {Lesage},
  \& {Richou}}]{majoqsart69}
{Matheron}, P., {Escarguel}, A., {Redon}, R., {Lesage}, A., \& {Richou}, J.
  2001, J.\, Quant.\,Spectrosc.\ Radiat.\ Transf.\, 69, 535

\bibitem[{{Mendoza} {et~al.}(1995){Mendoza}, {Eissner}, {LeDourneuf}, \&
  {Zeippen}}]{meadfoc28}
{Mendoza}, C., {Eissner}, W., {LeDourneuf}, M., \& {Zeippen}, C.~J. 1995, J.\
  Phys.\ B: At.\ Mol.\ Opt.\ Phys.\, 28, 3485

\bibitem[{{Nahar}(2002)}]{naadndt80}
{Nahar}, S.~N. 2002, At.~Data Nucl.~Data Tables, 80, 205

\bibitem[{{Nieva} \& {Przybilla}(2006)}]{niapjl639}
{Nieva}, M.~F., \& {Przybilla}, N. 2006, \apjl, 639, L39

\bibitem[{{Nieva} \& {Przybilla}(2007)}]{niaap467}
{Nieva}, M.~F., \& {Przybilla}, N. 2007, \aap, 467, 295

\bibitem[{{Nieva} \& {Przybilla}(2008)}]{niaap481}
{Nieva}, M.~F., \& {Przybilla}, N. 2008, \aap, 481, 199

\bibitem[{{Nussbaumer}(1986)}]{nuaap155}
{Nussbaumer}, H. 1986, \aap, 155, 205

\bibitem[{{Ochsenbein} {et~al.}(2000){Ochsenbein}, {Bauer}, \&
  {Marcout}}]{vizier}
{Ochsenbein}, F., {Bauer}, P., \& {Marcout}, J. 2000, \aaps, 143, 23

\bibitem[{{Odenkirchen} \& {Brosche}(1992)}]{odan313}
{Odenkirchen}, M., \& {Brosche}, P. 1992, Astron.~Nachr., 313, 69

\bibitem[{{Pfeiffer} {et~al.}(1998){Pfeiffer}, {Frank}, {Baumueller},
  {Fuhrmann}, \& {Gehren}}]{pfaaps130}
{Pfeiffer}, M.~J., {Frank}, C., {Baumueller}, D., {Fuhrmann}, K., \& {Gehren},
  T. 1998, \aaps, 130, 381

\bibitem[{{Poveda} {et~al.}(1967){Poveda}, {Ruiz}, \& {Allen}}]{pobdlotyt4}
{Poveda}, A., {Ruiz}, J., \& {Allen}, C. 1967, Bol.\ Obs.\
  Tonantzintla Tacubaya, 4, 86

\bibitem[{{Przybilla} {et~al.}(2006){Przybilla}, {Butler}, {Becker}, \&
  {Kudritzki}}]{praap445}
{Przybilla}, N., {Butler}, K., {Becker}, S.~R., \& {Kudritzki}, R.~P. 2006,
  \aap, 445, 1099

\bibitem[{{Przybilla} {et~al.}(2008{\natexlab{a}}){Przybilla}, {Nieva}, \&
  {Butler}}]{prapjl688}
{Przybilla}, N., {Nieva}, M.-F., \& {Butler}, K. 2008{\natexlab{a}}, \apjl,
  688, L103

\bibitem[{{Przybilla} {et~al.}(2008{\natexlab{b}}){Przybilla}, {Nieva},
  {Heber}, \& {Butler}}]{prapj684}
{Przybilla}, N., {Nieva}, M.~F., {Heber}, U., \& {Butler}, K.
  2008{\natexlab{b}}, \apjl, 684, L103

\bibitem[{{Ramspeck} {et~al.}(2001){Ramspeck}, {Heber}, \&
  {Edelmann}}]{raaap379}
{Ramspeck}, M., {Heber}, U., \& {Edelmann}, H. 2001, \aap, 379, 235

\bibitem[{{Schaller} {et~al.}(1992){Schaller}, {Schaerer}, {Meynet}, \&
  {Maeder}}]{scaaps96}
{Schaller}, G., {Schaerer}, D., {Meynet}, G., \& {Maeder}, A. 1992, \aaps, 96,
  269

\bibitem[{{Schlegel} {et~al.}(1998){Schlegel}, {Finkbeiner}, \&
  {Davis}}]{scapj500}
{Schlegel}, D.~J., {Finkbeiner}, D.~P., \& {Davis}, M. 1998, \apj, 500, 525

\bibitem[{{Tauris} \& {Takens}(1998)}]{taaap330}
{Tauris}, T.~M., \& {Takens}, R.~J. 1998, \aap, 330, 1047

\bibitem[{{Tenjes} {et~al.}(2001){Tenjes}, {Einasto}, {Maitzen}, \&
  {Zinnecker}}]{teaap369}
{Tenjes}, P., {Einasto}, J., {Maitzen}, H.~M., \& {Zinnecker}, H. 2001, \aap,
  369, 530

\bibitem[{{Tobin}(1985)}]{toaaps60}
{Tobin}, W. 1985, \aaps, 60, 459

\bibitem[{{Tobin}(1987)}]{tobin87}
{Tobin}, W. 1987, in IAU Colloq. 95, The Second Conference on Faint Blue Stars, ed. A.~G.~D.
  Philip, D.~S. Hayes, \& J.~W. Liebert (Schenectady: L. Davis Press), 149

\bibitem[{{van Leeuwen}(2007)}]{hipparcos}
{van Leeuwen}, F. 2007, Astrophysics and Space Science Library 350,
  \textit{Hipparcos}, the New Reduction of the Raw Data (Berlin: Springer)

\bibitem[{{Wiese} {et~al.}(1996){Wiese}, {Fuhr}, \& {Deters}}]{wijpcrd7}
{Wiese}, W.~L., {Fuhr}, J.~R., \& {Deters}, T.~M. 1996, J.\ Phys.\ Chem.\
  Ref.\ Data, Mon.\ 7 (Washington, DC: American Chemical Society and American
Institute of Physics)

\bibitem[{{Wiese} {et~al.}(1969){Wiese}, {Smith}, \& {Miles}}]{winsrds69}
{Wiese}, W.~L., {Smith}, M.~W., \& {Miles}, B.~M. 1969, Nat.\ Stand.\ Ref.\
  Data Ser., 22 (Washington, DC: Nat. Bur. Stand. [U.S.])

\bibitem[{{Xue} {et~al.}(2008){Xue}, {Rix}, {Zhao}, {Re Fiorentin}, {Naab},
  {Steinmetz}, {van den Bosch}, {Beers}, {Lee}, {Bell}, {Rockosi}, {Yanny},
  {Newberg}, {Wilhelm}, {Kang}, {Smith}, \& {Schneider}}]{xuapj684}
{Xue}, X.~X., {et~al.} 2008, \apj, 684, 1143

\end{thebibliography}

\clearpage
\LongTables
\begin{deluxetable*}{ccccccc}
\tablecolumns{7}
\tablewidth{0pc}  
\tablecaption{Spectral line analysis of \mbox{HIP\,60350} and \mbox{18 Peg}}
\label{linelist}
\tablehead{  
\colhead{$\lambda$(\AA)} & \colhead{$\chi(\rm{eV})$}   & \colhead{$\log gf$}    & \colhead{Accuracy} &
\colhead{Source}    & \multicolumn{2}{c}{$\log(\rm X/\rm H)+12$}\\
\cline{6-7}
\colhead{} & \colhead{}   & \colhead{}    & \colhead{} &
\colhead{}    & \colhead{\mbox{HIP\,60350}} & \colhead{\mbox{18 Peg}}}\\
\cline{1-7}
\startdata
C\,\textsc{ii}:	&					&				&			&			&						&					\\
3918.98			&	16.33			&	-0.533		&	B		&	WFD		&		$\ldots$		&	8.19			\\
3920.69			&	16.33			&	-0.232		&	B		&	WFD		&		$\ldots$		&	8.30			\\
4267.00			&	18.05			&	0.563		&	C+		&	WFD		&		8.69			&	8.19			\\
4267.26			&	18.05			&	0.716		&	C+		&	WFD		&		-''-			&	-''-			\\
4267.26			&	18.05			&	-0.584		&	C+		&	WFD		&		-''-			&	-''-			\\
5132.94			&	20.70			&	-0.211		&	B		&	WFD		&		8.93			&	8.37			\\
5133.28			&	20.70			&	-0.178		&	B		&	WFD		&		-''-			&	-''-			\\
5145.16			&	20.71			&	0.189		&	B		&	WFD		&		8.73			&	8.34			\\
5151.09			&	20.71			&	-0.179		&	B		&	WFD		&		$\ldots$		&	8.39			\\
5662.47			&	20.71			&	-0.249		&	B		&	WFD		&		$\ldots$		&	8.41			\\
6578.05			&	14.45			&	-0.087		&	C+		&	N02		&		$\ldots$		&	8.25			\\
6582.88			&	14.45			&	-0.388		&	C+		&	N02		&		$\ldots$		&	8.37			\\
6783.90			&	20.71			&	0.304		&	B		&	WFD		&		$\ldots$		&	8.44		\\[1mm]
				&					&				&			&			&		\textbf{8.79}	&	\textbf{8.33}	\\
				&					&				&			&			&	$\pm$\textbf{0.13}	& $\pm$\textbf{0.09}\\
				&					&				&			&			&						&					\\
N\,\textsc{ii}:	&					&				&			&			&						&					\\
3995.00   		&	18.50			&    0.163		&   B 		&     FFT   & 		  8.22  		&	  7.72			\\
4035.08			&   23.12 			&    0.599  	&	B      	&	  BB89  &		  $\ldots$  	&	  7.87			\\
4041.31	 		&	23.14    		&    0.748   	&	B   	&	  MAR   & 		  $\ldots$    	&	  7.90			\\
4176.16			& 	23.20   		&	 0.316   	&	B     	&	  MAR	&   	  $\ldots$    	&	  7.88			\\
4227.74	  		&	21.60   		&	-0.060   	&	B    	&	  WFD	& 		  $\ldots$   	&	  7.94			\\
4236.91   		&	23.24   		&	 0.383   	&	X    	&	  KB	& 		  $\ldots$   	&	  7.74			\\
4237.05   		&	23.24   		&	 0.553  	&	X    	&	  KB	& 		  $\ldots$   	&	  -''-			\\
4241.76  		&	23.24   		&	 0.210   	&	X    	&	  KB	& 		  $\ldots$   	&	  7.59			\\
4241.79	 		&	23.25   		&	 0.713   	&	X    	&	  KB	&  		  $\ldots$   	&	  -''-			\\
4447.03	 		&	20.41   		&	 0.221   	&	B    	&	  FFT	&  		  $\ldots$   	&	  7.85			\\
4601.48	  		&	18.47   		&	-0.452   	&	B+   	&	  FFT	&  			8.65   		&	  7.88			\\
4607.16	  		&	18.46   		&	-0.522   	&	B+   	&	  FFT	& 		  $\ldots$   	&	  7.74			\\
4630.54	  		&	18.48   		&	 0.080   	&	B+   	&	  FFT	&  		  $\ldots$   	&	  7.70			\\
4643.08   		&	18.48   		&	-0.371   	&	B+   	&	  FFT	& 		  $\ldots$  	&	  7.83			\\
5001.13   		&	20.65   		&	 0.257   	&	B    	&	  FFT	& 		  $\ldots$   	&	  7.91			\\
5001.48	  		&	20.65   		&	 0.435   	&	B    	&	  FFT	& 		  $\ldots$   	&	  -''-			\\
5005.15   		&	20.67   		&	 0.587   	&	B    	&	  FFT	&  		  $\ldots$   	&	  7.84			\\
5045.10	  		&	18.48   		&	-0.407   	&	B+   	&	  WFD	& 		  $\ldots$   	&	  7.68			\\
5666.63	  		&	18.47   		&	-0.104   	&	B+   	&	  MAR	& 		    7.97   		&	  7.67			\\
5676.02	  		&	18.46   		&	-0.356   	&	B+   	&	  MAR	&  		    7.96 		&	  7.66			\\
5679.56	  		&	18.48   		&	 0.221   	&	B+   	&	  MAR	&   		-''-  		&	  7.93			\\
				&					&				&			&			&		\textbf{8.20}	&	\textbf{7.80}	\\
				&					&				&			&			&	$\pm$\textbf{0.32}	& $\pm$\textbf{0.11}\\
				&					&				&			&			&						&					\\
O\,\textsc{i}:	&					&				&			&			&						&					\\
6155.96   		&	10.74   		&	-1.363  	&	B+   	&	  WFD   &		  	8.98 		&	   8.81			\\
6155.97   		&	10.74  			&	-1.011   	&	B+   	&	  WFD   &  			 -''- 		&	    -''-		\\
6155.99   		&	10.74   		&	-1.120   	&	B+   	&	  WFD   &			 -''- 		&	    -''-		\\
6156.74   		&	10.74   		&	-1.487   	&	B+   	&	  WFD   &			 -''-  		&		-''-		\\
6156.76   		&	10.74   		&	-0.898   	&	B+   	&	  WFD   &			 -''-    	&		-''-		\\
6156.78   		&	10.74   		&	-0.694   	&	B+   	&	  WFD   &   		 -''-		&	    -''-		\\
6158.15   		&	10.74   		&	-1.841   	&	B+   	&	  WFD   & 			 -''-    	&		8.81		\\
6158.17   		&	10.74   		&	-0.995   	&	B+   	&	  WFD   &  			 -''-     	&		-''-		\\
6158.19   		&	10.74   		&	-0.409   	&	B+   	&	  WFD   &  			 -''-     	&		-''-		\\
7771.94   		&	 9.15   		&	 0.354   	&	A    	&	  FFT   &  			 8.91    	&		8.54		\\
7774.17   		&	 9.15   		&	 0.207   	&	A    	&	  FFT   &  			 -''-     	&		8.75		\\
7775.39   		&	 9.15   		&	-0.015   	&	A    	&	  FFT   &  			 -''-     	&		8.68		\\
				&					&				&			&			&		\textbf{8.95}	&	\textbf{8.72}	\\
				&					&				&			&			&	$\pm$\textbf{0.05}	& $\pm$\textbf{0.11}\\
				&					&				&			&			&						&					\\
O\,\textsc{ii}:	&					&				&			&			&						&					\\
3911.96	  		&	25.66   		&	-0.014   	&	B+    	&	  FFT   & 	  $\ldots$    		&		 8.93		\\
3912.12   		&	25.66   		&	-0.907   	&	B+     	&	  FFT	& 	  $\ldots$   		&	 	 -''-		\\
4069.62   		&	25.63   		&	 0.144   	&	B+      &	  FFT	& 	  $\ldots$   		&		 8.81		\\
4069.88   		&	25.64   		&	 0.352   	&	B+  	&	  FFT	& 	  $\ldots$    		&		 -''-		\\
4072.72   		&	25.65   		&	 0.528   	&	B+   	&	  FFT	& 	  $\ldots$     		&		 8.88		\\
4075.86   		&	25.67   		&	 0.693   	&	B+    	&	  FFT	& 	  $\ldots$    		&		 8.94		\\
4185.45   		&	28.36   		&	 0.604   	&	D     	&	  WFD	&	  $\ldots$    		&		 8.75		\\
4317.14   		&	22.97   		&	-0.368   	&	B+   	&	  FFT	& 	  $\ldots$    		&		 8.86		\\
4319.63   		&	22.98   		&	-0.372   	&	B+   	&	  FFT	& 	  $\ldots$    		&		 8.69		\\
4366.89   		&	23.00   		&	-0.333   	&	B+    	&     FFT	& 	  $\ldots$    		&		 8.73		\\
4369.28   		&	26.23   		&	-0.383   	&	B+    	&	  FFT	&	  $\ldots$    		&		 8.88		\\
4414.90   		&	23.44   		&	 0.207   	&	B     	&	  FFT	& 	  $\ldots$    		&		 8.63		\\
4416.97   		&	23.42   		&	-0.043   	&	B    	&	  FFT	& 	  $\ldots$    		&		 8.95		\\
4452.38   		&	23.44   		&	-0.767   	&	B    	&	  FFT	&  	  $\ldots$    		&		 8.84		\\
4590.97   		&	25.66   		&	 0.331   	&	B+   	&	  FFT	& 	  $\ldots$   		&		 8.91		\\
4661.63   		&	22.98   		&	-0.269   	&	B+   	&	  FFT	& 	  $\ldots$    		&		 8.73		\\
4699.22   		&	26.23   		&	 0.238   	&	B    	&	  FFT	& 	  $\ldots$   		&		 8.88		\\
4705.35   		&	26.25   		&	 0.533   	&	B    	&	  FFT	& 	  $\ldots$    		&		 8.78		\\
				&					&				&			&			&	  $\ldots$			&	\textbf{8.82}	\\
				&					&				&			&			&	  $\ldots$			& $\pm$\textbf{0.10}\\
				&					&				&			&			&						&					\\
Mg\,\textsc{ii}:&					&				&			&			&						&					\\
4384.64 		&  10.00   			&	-0.792  	&	B+     	&	  FW    &	  $\ldots$   		&		 7.56		\\
4390.51  		&  10.00   			&	-1.703   	&	D    	&	  FW    &	  $\ldots$    		&		 7.54		\\
4390.57  		&  10.00   			&	-0.530  	&	B+   	&	  FW    &	  $\ldots$    		&		 -''-		\\
4433.99			&  10.00   			&	-0.900  	&	C+   	&	  FW    &	  $\ldots$    		&		 7.53		\\
4481.13  		&   8.86   			&	 0.730   	&	B+   	&	  FW    &   	7.38   			&		 7.41		\\
4481.15    		&   8.86   			&	-0.570   	&	B+   	&	  FW    &       -''-    		&		 -''-		\\
4481.33    		&   8.86   			&	 0.575   	&	B+   	&	  FW    &       -''-     		&		 -''-		\\
				&					&				&			&			&		\textbf{7.38}	&	\textbf{7.51}	\\
				&					&				&			&			&		$\ldots$		& $\pm$\textbf{0.07}\\
				&					&				&			&			&						&					\\
Si\,\textsc{ii}:&					&				&			&			&						&					\\
4621.42 		&   12.53 			&   -0.608  	&	D     	&	  MELZ  & 		  $\ldots$   	&	  	7.55		\\
4621.70  		&   12.53  			&   -1.754   	&	E    	&	  MELZ  &		  $\ldots$    	&		-''-		\\
4621.72 		&   12.53  			&   -0.453   	&	D    	&	  MELZ  & 		  $\ldots$    	&		-''-		\\
5041.02			&   10.07   		&	 0.029   	&	B     	&	  MERLR &  			 7.72    	&		7.34		\\
5055.98			&   10.07    		&	 0.523   	&	B     	&	  MERLR &		  $\ldots$   	&		7.43		\\
5056.32			&   10.07   		&   -0.492   	&	B    	&	  MERLR &		  $\ldots$   	&		-''-		\\
5185.54			&   12.84  			&   -0.059   	&	D    	&	  MELZ  &  		  $\ldots$    	&		7.59		\\
5978.93			&   10.07   		&	 -0.06    	&	D    	&	  WSM   & 		  $\ldots$    	&		7.35		\\
6347.11			&    8.12   		&	 0.177   	&	B+   	&	  FFTI  & 		  	 7.71  		&		7.61		\\
6371.37			&    8.12   		&   -0.126   	&	B    	&	  FFTI  & 		 	 7.63  		&		7.44		\\
				&					&				&			&			&	\textbf{7.69}		&	\textbf{7.47}	\\
				&					&				&			&			&	$\pm$\textbf{0.05}	& $\pm$\textbf{0.11}\\
				&					&				&			&			&						&					\\
Si\,\textsc{iii}:&					&				&			&			&						&					\\
4552.62  		&   19.02   		&	 0.292   	&   B+   	&     FFTI  & 		    $\ldots$    &   	7.61		\\
4567.84			&   19.02   		&	 0.068   	&   B+   	&     FFTI  & 		    7.52    	&    	7.48		\\
4574.76			&   19.02   		&	-0.409   	&   B    	&     FFTI  & 		    7.73    	&    	7.49		\\
4813.33			&   25.98   		&	 0.708   	&   B+   	&     BMZ   & 		    $\ldots$   	&     	7.61		\\
5739.73			&   19.72   		&	-0.096   	&   B+   	&     FFTI,N86  & 		$\ldots$   	&     	7.64		\\
				&					&				&			&			&	\textbf{7.62}		&	\textbf{7.57}	\\
				&					&				&			&			&	$\pm$\textbf{0.15}	& $\pm$\textbf{0.08}\\
				&					&				&			&			&						&					\\
S\,\textsc{ii}: &					&				&			&			&						&					\\
3990.91			&   15.90   		&   -0.30   	&    E   	&     WSM	&  		    $\ldots$	&     7.24			\\
3998.76			&   15.87   		&    0.06   	&    E   	&      FW	&       	$\ldots$	&     7.17			\\
4028.75			&   15.90   		&   -0.00   	&    D   	&      FW	&	        $\ldots$	&     7.20			\\
4032.76			&   16.25   		&    0.24   	&    D   	&      FW	&   	    $\ldots$	&     7.07			\\
4153.06			&   15.90   		&    0.62   	&    D   	&      FW	&	  		7.49		&     7.08			\\
4162.66			&   15.94   		&    0.78   	&    D   	&      FW	&	   		7.31		&     6.98			\\
4168.38			&   15.87   		&   -0.16   	&    E   	&      WSM	&     		$\ldots$	&     7.10			\\
4189.68			&   15.90   		&   -0.05   	&    E   	&      FW 	&      		$\ldots$	&     7.13			\\
4217.18			&   15.94   		&   -0.15   	&    E   	&      WSM	&      		$\ldots$	&     7.07			\\
4217.30			&   14.85   		&   -1.479   	&    B   	&      FFTI	&     		$\ldots$	&     -''-			\\
4267.76			&   16.10   		&    0.29   	&    E   	&      FW  	&    		$\ldots$	&     7.19			\\
4269.72			&   16.09   		&   -0.12   	&    E   	&      WSM 	&     		$\ldots$	&     7.11			\\
4278.50			&   16.09   		&   -0.11   	&    E   	&      FW  	&     		$\ldots$	&     7.07			\\
4282.59			&   16.10   		&   -0.01   	&    E   	&      WSM 	&     		$\ldots$	&     7.16			\\
4294.40			&   16.13   		&    0.58   	&    D   	&      FW  	&     		$\ldots$	&     6.95			\\
4318.64			&   16.13   		&   -0.08   	&    E   	&      WSM 	&     		$\ldots$	&     7.22			\\
4524.67			&   15.07   		&   -0.744   	&    B   	&      FFTI	&     		$\ldots$	&     6.92			\\
4524.94			&   15.07   		&    0.032   	&    B   	&      FFTI	&     		$\ldots$	&     6.92			\\
4656.75			&   13.58   		&   -0.827   	&    B   	&      FFTI	&     		$\ldots$	&     7.25			\\
4716.27			&   13.62   		&   -0.366   	&    B   	&      FFTI	&     		$\ldots$	&     7.16			\\
4815.55			&   13.67   		&    0.068   	&    B   	&      FFTI	&     		7.56		&     6.99			\\
4917.19			&   14.00   		&   -0.375   	&    B   	&      FFTI	&     		$\ldots$	&     7.24			\\
4942.47			&   13.58   		&   -1.034   	&    B   	&      FFTI	&     		$\ldots$	&     7.15			\\
5009.56			&   13.62   		&   -0.235   	&    B   	&      FFTI	&     		$\ldots$	&     7.11			\\
5014.04			&   14.07   		&    0.046   	&    B   	&      FFTI	&     		$\ldots$	&     7.12			\\
5027.20			&   13.09   		&   -0.664   	&    B   	&      FFTI	&     		$\ldots$	&     7.07			\\
5032.43			&   13.67   		&    0.187   	&    B   	&      FFTI	&	  		7.15		&     6.93			\\
5142.32			&   13.15   		&   -1.004   	&    B   	&      FFTI	&     		$\ldots$	&     7.00			\\
5201.03			&   15.07   		&    0.088   	&    B   	&      FFTI	&	  		7.55		&     6.97			\\
5201.37			&   15.07   		&   -0.712   	&    B   	&      FFTI	&	  		 -''-		&     -''-			\\
5212.27			&   15.07   		&   -1.444   	&    B   	&      FFTI	&     		$\ldots$	&     6.99			\\
5212.62			&   15.07   		&    0.316   	&    B   	&      FFTI	&     		$\ldots$	&     -''-			\\
5320.72			&   15.07   		&    0.431   	&    B   	&      FFTI	&     		$\ldots$	&     6.96			\\
5345.71			&   15.07   		&    0.289   	&    B   	&      FFTI	&	  		7.22		&     6.98			\\
5346.08			&   15.07   		&   -1.161   	&    B   	&      FFTI	&	  		-''-		&     6.96			\\
5428.65			&   13.58   		&   -0.175   	&    B   	&      FFTI	&    		$\ldots$	&     7.03			\\
5432.79			&   13.62   		&    0.205   	&    B   	&      FFTI	&     		$\ldots$	&     7.01			\\
5453.85			&   13.67   		&    0.442   	&    B   	&      FFTI	&     		$\ldots$	&     6.92			\\
5473.61			&   13.58   		&   -0.226   	&    B   	&      FFTI	&     		$\ldots$	&     7.05			\\
5509.70			&   13.62   		&   -0.175   	&    B   	&      FFTI	&     		7.3			&     7.08			\\
5526.24			&   13.70   		&   -0.871   	&    B   	&      FFTI	&     		$\ldots$	&     6.90			\\
5564.95			&   13.67   		&   -0.335   	&    B   	&      FFTI	&     		$\ldots$	&     7.09			\\
5578.87			&   13.68   		&   -0.727   	&    B   	&      FFTI	&     		$\ldots$	&     7.10			\\
5606.15			&   13.73   		&    0.123   	&    B   	&      FFTI	&	  		7.44		&     6.89			\\
5616.63			&   13.66   		&   -0.845   	&    B   	&      FFTI	&    		$\ldots$	&     7.19			\\
5647.02			&   14.00   		&    0.021   	&    B   	&      FFTI	&	   		7.56		&     7.20			\\
5660.00			&   13.68   		&   -0.220   	&    B   	&      FFTI	&	   		7.23		&     7.02			\\
5664.77			&   13.66   		&   -0.426   	&    B   	&      FFTI	&	  		 -''-		&     7.11			\\
5819.25			&   14.07   		&   -0.713   	&    B   	&      FFTI	&    		$\ldots$	&     7.26			\\
6286.34			&   14.16   		&   -0.750   	&    B   	&      FFTI	&     		$\ldots$	&     7.11			\\
6286.94			&   14.29   		&   -0.157   	&    B   	&      FFTI	&     		$\ldots$	&     -''-			\\
6384.89			&   14.16   		&   -0.695   	&    B   	&      FFTI	&     		$\ldots$	&     7.29			\\
6397.35			&   14.16   		&   -0.434   	&    B   	&      FFTI	&     		$\ldots$	&     7.25			\\
				&					&				&			&			&	\textbf{7.38}		&	\textbf{7.08}	\\
				&					&				&			&			&	$\pm$\textbf{0.16}	& $\pm$\textbf{0.11}\\
				&					&				&			&			&						&					\\
S\,\textsc{iii}:&					&				&			&			&						&					\\
4253.50			&   18.24			&    0.095		&    B    	&	   FFTI &		    $\ldots$	&     7.02			\\
				&					&				&			&			&						&					\\
Fe\,\textsc{ii}:&					&				&			&			&						&					\\
4178.86			&    2.58   		&   -2.48   	&    C    	&      FMW	&      $\ldots$ 		&     7.59			\\
4233.17			&    2.58   		&   -2.00   	&    C    	&      FMW	&      $\ldots$ 		&     7.62			\\
4303.17			&    2.70   		&   -2.49   	&    C    	&      FMW	&      $\ldots$ 		&     7.59			\\
4351.76			&    2.70   		&   -2.10   	&    C    	&      FMW	&      $\ldots$ 		&     7.44			\\
4508.28			&    2.86   		&   -2.31   	&    D    	&      KB 	&      $\ldots$ 		&     7.52			\\
4515.34			&    2.84   		&   -2.48   	&    D    	&      FMW	&      $\ldots$ 		&     7.50			\\
4520.22			&    2.81   		&   -2.60   	&    D    	&      FMW	&      $\ldots$ 		&     7.56			\\
4522.63			&    2.84   		&   -2.11   	&    C    	&      KB 	&      $\ldots$ 		&     7.50			\\
4549.47			&    2.82   		&   -1.75   	&    C    	&      FMW	&      $\ldots$ 		&     7.49			\\
4555.75			&    2.83   		&   -2.32   	&    D    	&      KB 	&      $\ldots$ 		&     7.56			\\
4576.34			&    2.84   		&   -3.04   	&    D    	&      FMW	&      $\ldots$ 		&     7.59			\\
4582.83			&    2.84   		&   -3.10   	&    C    	&      FMW	&      $\ldots$ 		&     7.63			\\
4583.83			&    2.81   		&   -2.02   	&    D    	&      FMW	&      $\ldots$ 		&     7.65			\\  
4629.33			&    2.81   		&   -2.37   	&    D    	&      FMW	&      $\ldots$ 		&     7.53			\\
4923.92			&    2.89   		&   -1.32   	&    C    	&      FMW	&      $\ldots$ 		&     7.58			\\
5018.44			&    2.89   		&   -1.22   	&    C    	&      FMW	&      7.18     		&	  7.42			\\
5197.57			&    3.23   		&   -2.23   	&    C    	&      KB 	&      $\ldots$ 		&     7.63			\\
5234.62			&    3.22   		&   -2.15   	&    C    	&      KB 	&      $\ldots$ 		&     7.67			\\
5316.61			&    3.15   		&   -1.85   	&    C    	&      FMW	&      $\ldots$ 		&     7.62			\\
				&					&				&			&			&	\textbf{7.18}		&	\textbf{7.56}	\\
				&					&				&			&			&	 $\ldots$			& $\pm$\textbf{0.07}\\
				&					&				&			&			&						&					\\
Fe\,\textsc{iii}:&					&				&			&			&						&					\\
4352.57			&    8.25   		&  -2.870   	&   X    	&      KB   &       $\ldots$ 		&     7.48			\\
4395.75			&    8.26   		&  -2.595   	&   X    	&      KB   &       $\ldots$ 		&     7.52			\\
4431.02			&    8.25   		&  -2.572   	&   X    	&      KB   &       $\ldots$ 		&     7.50			\\
5073.90			&    8.65   		&  -2.557   	&   X    	&      KB   &       $\ldots$ 		&     7.52			\\
5086.70			&    8.66   		&  -2.590   	&   X    	&      KB   &       $\ldots$ 		&     7.59			\\
5127.39			&    8.66   		&  -2.218   	&   X    	&      KB   &  		$\ldots$ 		&     7.43			\\
5127.63			&    8.66   		&  -2.564   	&   X    	&      KB   &       $\ldots$ 		&     -''-			\\
5156.11			&    8.64   		&  -2.018   	&   X    	&      KB   &       $\ldots$ 		&     7.41			\\
				&					&				&			&			&		$\ldots$		&	\textbf{7.49}	\\
				&					&				&			&			&		$\ldots$		& $\pm$\textbf{0.06}\\
\cline{1-7}
\enddata
\end{deluxetable*}
\end{document}